# Polaron Masses in CH$_3$NH$_3$PbX$_3$ Perovskites Determined by Landau Level Spectroscopy in Low Magnetic Fields

Yasuhiro Yamada,[1,*] Hirofumi Mino,[2] Takuya Kawahara,[3] Kenichi Oto,[1] Hidekatsu Suzuura,[4] and Yoshihiko Kanemitsu[5,**]

[1] *Graduate School of Science, Chiba University, Inage, Chiba 263-8522, Japan*

[2] *College of Liberal Arts and Sciences, Chiba University, Inage, Chiba 263-8522, Japan*

[3] *Graduate School of Education, Chiba University, Inage, Chiba 263-8522, Japan*

[4] *Graduate School of Engineering, Hokkaido University, Sapporo 060-8628, Japan*

[5] *Institute for Chemical Research, Kyoto University, Uji, Kyoto 611-0011, Japan*

## Abstract

We investigate the electron–phonon coupling in CH$_3$NH$_3$PbX$_3$ lead halide perovskites through the observation of Landau levels and high-order excitons at weak magnetic fields, where the cyclotron energy is significantly smaller than the longitudinal optical phonon energy. The reduced masses of the carriers and the exciton binding energies obtained from these data are clearly influenced by polaron formation. We analyze the field-dependent polaronic and excitonic properties, and show that they can be quantitatively reproduced by the Fröhlich large polaron model.

Corresponding authors: * yasuyamada@chiba-u.jp, ** kanemitu@scl.kyoto-u.ac.jp

Organic–inorganic hybrid lead halide perovskites $CH_3NH_3PbX_3$ [X = I, Br, and Cl] are a novel class of high-quality semiconductor materials that are considered suitable for use in solar cells [1,2], light-emitting devices [3,4], sensors [5], and other optoelectronic devices [6]. These materials may constitute the basis for future semiconductor physics and innovative devices beyond conventional semiconductors. In particular, the lead halide perovskites combine features such as high photoluminescence (PL) quantum yields and the capability to tune the bandgap over a wide energy range [7–14]. Therefore, much effort has been devoted to the clarification of the physical origin of the outstanding performance of this class of materials. For example, a large electron–phonon coupling constant and the resulting polaron formation have been considered to dominate the unique features of halide perovskites [15–19]. The strength of the conventional large polaron effect can be characterized using the Fröhlich coupling constant, and it has been shown that the Fröhlich coupling constants of halide perovskites are much larger than those of conventional inorganic semiconductors (*e.g.* $\approx 2$ for $CH_3NH_3PbBr_3$ but only 0.07 for GaAs and 0.29 for CdTe). Furthermore, the soft lattice nature and the cation dipole orientation have also been considered as possible origins of a pronounced polaron effect [15,16,19]. Several research groups have considered a large polaron formation induced by the conventional Fröhlich mechanism or a ferroelectric-like dielectric response [15,16,19], and a small polaron formation coupled to local structural lattice distortions [17,18]. However, the polaron formation mechanisms in halide perovskites still remain unclear. To clarify the actual importance of the polaron effect, it is essential to determine the carrier mass, which is influenced by polaron effect.

Magneto-optical spectroscopy is the most powerful methods for deriving the band-edge optical parameters of semiconductors, including the carrier reduced masses and the exciton binding energy. Previous magneto-optical studies on halide perovskites estimated the reduced mass by employing extremely strong pulsed magnetic fields up to 130 T [20,21], which enables the study of reduced mass through the observation of Landau levels even in samples with low crystalline quality. However, a careful interpretation is required for materials with electron–phonon coupling. The impact of polaron effect on the optical properties is dependent on the applied magnetic field [22]. This means that the electron–phonon interaction contributes differently to the optical properties under strong and weak magnetic fields. Consequently, it is essential to measure high-quality crystals under weak magnetic fields in order to accurately estimate the band-edge optical properties in the low-field limit, which is more relevant to the actual device operation and material research of halide perovskites.

In this Letter, we report magnetoreflectance spectroscopy of $CH_3NH_3PbX_3$ bulk crystals at low temperatures. We employ relatively weak magnetic fields up to 7 T, which corresponds to a regime where the cyclotron energy $\hbar\omega_c$ is smaller than the exciton Rydberg energy. Under such weak magnetic fields, the LO phonon energies of $CH_3NH_3PbI_3$ and $CH_3NH_3PbBr_3$ are larger than $\hbar\omega_c$, and thus the phonon screening effect plays an essential role. From the observation of Landau levels and higher-order exciton Rydberg states at weak fields, we are able to accurately estimate the reduced masses and the exciton binding energies in $CH_3NH_3PbI_3$ and $CH_3NH_3PbBr_3$. For both materials, we find that the reduced mass at low magnetic fields differs from that at high fields. The obtained data set is consistent and allows us to assess the mass enhancement by polaron effect, which proves a relatively small contribution of exciton-phonon interaction to the carrier mass of halide perovskites.

For the optical measurements, we fabricated $CH_3NH_3PbX_3$ [X = I, Br, and Cl] bulk crystals and cleaved them to obtain clean surfaces. We conducted polarization-resolved reflectance measurements at 1.5 K using a Helmholtz superconducting magnet in the Faraday configuration. The detailed experimental setup and crystal growth technique are explained in the Supplemental Material [23].

The reflectance spectra of the $CH_3NH_3PbI_3$ bulk crystal at 1.5 K under different magnetic field are shown in Fig. 1, where the results for excitation with left ($\sigma^+$) and right circularly polarized ($\sigma^-$) light are shown in Figs. 1(a) and (b), respectively. The spectrum in the case without applied magnetic field (0 T) exhibits an inflection point at approximately 1.637 eV, which corresponds to the 1s exciton resonance energy. The broadening of the 1s exciton (*i.e.*, the energy separation between the positive and negative peaks) is 1.3 meV, which is much narrower than the values that have been reported previously [20,35]. This indicates that the crystal quality of our sample is relatively high.

Both Figs. 1(a) and (b) clarify that the 1s exciton resonance slightly shifts to higher energies as the magnetic field increases. In addition, small peaks appear at the high-energy side as well as a peculiar feature immediately below the 1s exciton resonance in the case of $\sigma^+$ excitation. The latter feature is discussed in the Supplemental Material [23]. The magnified views of the high-energy peaks in the data for 7 T are shown in the right upper corners of the figures. The dependences of their peak energies on the magnetic field are summarized in Fig. 1(c).

Because the peak energies of Landau levels exhibit an almost linear dependence on the magnetic field [20,36,37], we assigned the peaks in the range from 1.645 to 1.655 eV to the Landau levels. The magnetic-field dependences of the Landau levels can be fitted to the equation $E_g + E_N(B) \pm \frac{1}{2} g_{\text{eff}} \mu_B B$. Here, $E_g$ denotes the bandgap energy, $E_N$ is the energy of the $N$-th Landau level, $N = 0$, 1, 2,… is the quantum number; $g_{\text{eff}}$, $\mu_B$, and $B$ are the effective g-factor, the Bohr magneton, and the magnetic field, respectively. We determined $g_{\text{eff}} = 1.68$ ($\pm 0.01$) from the results. If there was no electron–phonon coupling, $E_N = \left(N + \frac{1}{2}\right)\hbar\omega_c$ would hold, where $\omega_c = \frac{eB}{\mu^b}$ ($\mu^b$ is the reduced band mass). In other words, the Landau levels would depend linearly on the magnetic field. However, in the presence of electron–phonon coupling, the Landau energy slightly deviates from the above-mentioned linear dependence as a result of a mixing of Landau levels and LO phonons [36,37]. To proceed, we use $\omega_{LO}$ to denote the single effective LO phonon frequency (see Supplemental Material [23]) and define the polaron cyclotron frequency $\omega_c^*$ by

$$\hbar\omega_c^*(B) = E_2(B) - E_1(B). \tag{1}$$

The reduced polaron mass is then given by $\mu^{pol}(B) = eB/\omega_c^*(B)$, and the experimentally determined field dependence of $\mu^{pol}$ is plotted in the inset of Fig. 1(c) (see the open and closed red squares). The obtained $\mu^{pol}$ values lie in the range $0.14$– $0.16\ m_0$ ($m_0$: electron rest mass) for magnetic fields between 4 and 7 T. Compared to the values reported in high-field experiments ($\mu \sim 0.10\ m_0$) [20], the values obtained in our experiment are larger.

We consider that the above-mentioned difference is a result of the magnetic-field-dependent screening by LO phonons (see Supplemental Material [23]); the reduced mass estimated from the Landau levels at high magnetic fields ($\omega_c \gg \omega_{LO}$) reflects the bare carrier mass without electron–phonon interactions. The mass enhancement factor at zero magnetic field is approximately $1 + \alpha_e/6 + 0.0236\alpha_e^2$, where $\alpha_e$ is the Fröhlich coupling constant for electrons and given by [38]

$$\alpha_e = (\varepsilon_\infty^{-1} - \varepsilon_0^{-1})\sqrt{m_e^b e^4 / 2\hbar^3 \omega_{LO}}. \qquad (2)$$

Here, $\varepsilon_\infty$ and $\varepsilon_0$ are the low- and high-frequency dielectric constants, respectively, and $m_e^b$ denotes the electron band mass for a rigid lattice. The Fröhlich coupling constant for holes is obtained by replacing the index $e$ with $h$. By using the parameters provided in Table 1, we obtain $\alpha_e = 1.42$ and $\alpha_h = 1.67$. These values are significantly larger than unity, indicating that the bare carrier mass determined at high magnetic fields should be significantly smaller than the carrier mass determined at 0 T. A similar result has also been reported for ZnSe, where a lighter effective mass appears at high magnetic fields with $\hbar\omega_c \gg \hbar\omega_{LO}$ [39]. Furthermore, the inset of Fig. 1(c) shows that $\mu^{pol}$ decreases as the magnetic field decreases. Such a magnetic-field dependence of $\mu^{pol}$ evidences the presence of mixing between Landau levels and LO phonons, and is a fingerprint of a polaron cyclotron motion [37-39]. The dashed curve in the inset of Fig. 1(c) corresponds to the theoretical prediction for $\alpha_e = \alpha_h = 1.4$. These values for $\alpha$ were obtained by interpolating the data presented in Ref. [37]. These results evidence that the Fröhlich model can describe the experimental results well, and therefore our estimated value for $\mu^{pol}$ at 0 T is 0.13 $m_0$. Note that the estimated coupling constant is consistent with the previous first-principle calculation ($\alpha = 1.4$) [40].

In Fig. 1(c), the 1s exciton resonance energies are also plotted for both excitation conditions (see the open and closed circles). The magnetic-field dependence of the 1s exciton energy obeys the function $C_0 B^2 \pm \frac{1}{2} g_{ex} \mu_B B$, where $C_0$ and $g_{ex}$ are the diamagnetic constant and exciton g-factor, respectively. By using a global fitting procedure that accounts for both $\sigma^+$ and $\sigma^-$ excitation conditions, we estimated $C_0 = 15.9 \ (\pm 0.1) \ \mu eV/T^2$ and $g_{ex} = 1.63$, which is consistent with the effective g-factor estimated from the Landau levels. The exciton binding energy can be roughly estimated from the linear extrapolation of the Landau energies to 0 T as shown in Fig. 1(c), resulting in about 3.1 ($\pm 0.5$) meV, which is quite different from the binding energies obtained by using the Wannier–Mott exciton model for the regime $\hbar\omega_c \gg \hbar\omega_{LO}$ (*cf.* $R^* = 16$ meV has been reported in thin films [20] and bulk crystals [21]). We consider that this difference can be attributed to the difficulty in estimating zero-field parameters from a simple extrapolation using high-field data.

To further understand the polaron effect in halide perovskites, it is important to investigate the dependence of the excitonic properties on the halogen. The reflectance spectra of the $CH_3NH_3PbBr_3$ bulk crystal at 0 and 7 T are shown in Fig. 2(a). The 1s exciton resonance appears at 2.254 eV. By applying a magnetic field, the exciton resonance energy shifts due to the Zeeman splitting. The shift is relatively small and the degree of the shift depends on the excitation polarization. We observed a linear dependence on the magnetic field and estimated a g-factor $g_{ex}$ of 2.30 ($\pm 0.01$). As indicated by the arrow, a small dip appears below the 1s exciton energy when a magnetic field is applied. Similar to the case of $CH_3NH_3PbI_3$, this spectral feature was only observed when we used the $\sigma^+$ excitation condition.

To accurately determine the exciton energy structure, we investigated the differences between the reflectance spectra obtained using $\sigma^+$ and $\sigma^-$ excitation light. We define the degree of circular polarization $p$ as $p = (I^+ - I^-)/(I^+ + I^-)$, where $I^+$ and $I^-$ are the reflectance intensities obtained for $\sigma^+$ and $\sigma^-$ excitation, respectively. Note that, when the Zeeman splitting is small, the circular polarization spectrum $p$ corresponds to the first derivative of the reflectance spectrum. This enables us to resolve small structures in the reflectance spectrum, such as higher-order exciton resonances.

Figure 2(b) shows the circular polarization spectra of $CH_3NH_3PbBr_3$ for different magnetic field. We confirmed a strong peak at 2.254 eV, which agrees well with previous reports and thus was assigned to the 1s free exciton resonance [41,42]. The 1s peak energy of $CH_3NH_3PbBr_3$ exhibits an extremely small diamagnetic shift compared with that of $CH_3NH_3PbI_3$, and the estimated diamagnetic coefficient is $C_0 = 2.86 (\pm 0.1)$ μeV/T² [23]. Above the 1s peak energy, a small peak appears at approximately 2.265 eV, which exhibits a blueshift as the magnetic field increases. Based on the magnetic-field dependence of the resonance energy, we assigned this peak to the 2s exciton state [23]. A magnification of the data at 7 T in the range between 2.263 and 2.286 eV is shown in the inset of Fig. 2(b). Above 2.27 eV, we can observe three peaks where the left and the right peaks have almost the same energy separation from the center peak, as indicated by the dotted lines in the inset of Fig. 2(b). Therefore, we consider that these peaks correspond to Landau levels.

Figure 2(c) summarizes the 1s and 2s exciton states and the Landau levels of $CH_3NH_3PbBr_3$ as a function of the magnetic field. Similar to the case of $CH_3NH_3PbI_3$, we can determine the magnetic-field dependence of $\mu^{pol}$, and the result is shown in Fig. 2(d). The broken curve corresponds to the theoretical prediction for $\alpha_e = \alpha_h = 2.5$, which well reproduces the experimentally obtained trend. The value for $\alpha$ was obtained by interpolating the data presented in Ref. [37] and is consistent with the values calculated from Eq. (2) ($\alpha_e = 2.00$ and $\alpha_h = 2.24$; see Table 1). Based on this result, we determined $\mu^{pol}(0) = 0.20(\pm 0.01)\ m_0$ for $CH_3NH_3PbBr_3$.

The estimated 1s and 2s exciton energies of $CH_3NH_3PbBr_3$ at 0 T are 2.254 and 2.264 eV, respectively. Moreover, a bandgap energy of 2.265 eV can be roughly estimated by the linear extrapolation of the first Landau level to 0 T. Hence, a 1s exciton binding energy of 11 ($\pm 1$) meV was determined. This value is smaller than the previously reported exciton binding energies of $CH_3NH_3PbBr_3$ ($\approx 15.3$ meV [41] and 25 meV [21]), which have been estimated by using the Wannier–Mott model. Because the bandgap energy predicted by the Wannier–Mott model (2.268 eV) exceeds the first Landau level as long as $B < 5$ T, we consider that the Wannier–Mott model is not suitable to explain our experimental results. Moreover, the effective dielectric constant calculated under the assumption of the Wannier–Mott model is $\varepsilon_{eff} = 17$, which is nearly at the middle point between the high-frequency (5.2) and low-frequency (28) values [43]. Such a situation is typically observed in the materials with electron–phonon coupling, which results in a deviation of the high-order exciton resonance energies from the Wannier–Mott model [28].

Before we discuss the significance of the above results for applications, we quantitatively calculate the polaronic and excitonic material parameters based on the Haken model [28], which provides the effective dielectric constant for excitons depending on the ratio of the electron–hole distance to the polaron radius. The polaron radii for electrons and holes, $r_e^p$ and $r_h^p$, were calculated using $r^p = (\hbar/2m^b\omega_{LO})^{1/2}$ [28,30]. The literature values used in the calculation are listed in Table 1. To consider the multiple-LO phonons in the orthorhombic phase, we adopt a single effective LO phonon energy, which corresponds to the weighted average of multiple LO phonons [23,32,33]. More details and the data of the $CH_3NH_3PbCl_3$ bulk crystal are provided in Supplemental Material [23]. The results are summarized in Table 1. The calculated exciton binding energies and the values for $\mu^{pol}$ agree well with our experimental results, indicating that the Fröhlich-type electron–phonon coupling dominates the band-edge optical properties.

The above results suggest the following polaron physics in halide perovskites: Except the Fröhlich-type electron–phonon coupling, there are no significant contributions from other polaron formation mechanisms, such as small polarons and ferroelectric-like large polarons. Even more significant is the fact that the estimated polaron mass enhancement at 0 T (compared to the bare carrier mass given in Table 1) is too small to account for the actually observed moderate carrier mobility of halide perovskites. In other words, our results suggest that the intrinsic carrier mobility values should be larger than those reported so far. Therefore, we need to consider the reason why the reported experimental and theoretical carrier mobility values vary over a wide range [15,44–50]. The actually observed carrier mobility may be relatively low due to a difficulty in the doping method, an enhanced carrier scattering by shallow traps, and grain/domain boundaries. The carrier transport properties of halide perovskites should also be revisited.

In summary, we have presented reflection spectra of $CH_3NH_3PbX_3$ under weak magnetic fields. By changing the halogen from I to Br to Cl, the electron-phonon coupling becomes stronger, which results in the larger mass enhancement caused by the polaron effect. We found that the reduced masses are significantly larger than those estimated from high-field experiments. Furthermore, we discovered a deviation of the exciton energies from the simple Wannier–Mott model in $CH_3NH_3PbBr_3$. The calculation based on the Haken model revealed that the excitonic properties are determined by the phonon screening of electron–hole Coulomb interactions. The difference in the LO phonon energy can account for the halogen dependence of the exciton binding energy.

**Acknowledgments**

Part of this work was supported by JST-CREST (Grant No. JPMJCR16N3) and JSPS KAKENHI (Grant No. JP19K03683, JP19H05465, JP20K03798). The authors would like to thank S. Sakaguchi for the atomic force microscopy measurements.

**References**

[1] A. Kojima, K. Teshima, Y. Shirai, and T. Miyasaka, J. Am. Chem. Soc. **131**, 6050 (2009).

[2] M. A. Green, E. D. Dunlop, J. Hohl‐Ebinger, M. Yoshita, N. Kopidakis, A. W.Y. Ho‐Baillie, Prog. Photovolt: Res. Appl. **27**, 3 (2019).

[3] Z.-K. Tan, R. S. Moghaddam, M. L. Lai, P. Docampo, R. Higler, F. Deschler, M. Price, A. Sadhanala, L. M. Pazos, D. Credgington, F. Hanusch, T. Bein, H. J. Snaith, and R. H. Friend, Nat. Nanotech. **9**, 687 (2014).

[4] A. Sadhanala, S. Ahmad, B. Zhao, N. Giesbrecht, P. M. Pearce, F. Deschler, R. L. Z. Hoye, K. C. Gödel, T. Bein, P. Docampo, S. E. Dutton, M. F. L. De Volder, and R. H. Friend, Nano Lett. **15**, 6095 (2015).

[5] L. Gu, M. M. Tavakoli, D. Zhang, Q. Zhang, A. Waleed, Y. Xiao, K.‐H. Tsui, Y. Lin, L. Liao, J. Wang, and Z. Fan, Adv. Mater. **28**, 9713 (2016).

[6] T. Handa, H. Tahara, T. Aharen, and Y. Kanemitsu, Sci. Adv. **5**, eaax0786 (2019).

[7] M. R. Filip, G. E. Eperon, H. J. Snaith, and F. Giustino, Nat. Commun. **5**, 5757 (2014).

[8] L. N. Quan, B. P. Rand, R. H. Friend, S. G. Mhaisalkar, T.-W. Lee, and E. H. Sargent. Chem. Rev. **119**, 7444 (2019).

[9] F. Deschler, M. Price, S. Pathak, L. E. Klintberg, D.-D. Jarausch, R. Higler, S. Hüttner, T. Leijtens, S. D. Stranks, H. J. Snaith, M. Atatüre, R. T. Phillips, and R. H. Friend, J. Phys. Chem. Lett, **5**, 1421 (2014).

[10] K. Kojima, K. Ikemura, K. Matsumori, Y. Yamada, Y. Kanemitsu, and S. F. Chichibu APL Materials **7**, 071116 (2019).

[11] S. T. Ha, C. Shen, J. Zhang, and Q. Xiong, Nat. Photon. **10**, 115 (2016).

[12] Y. Yamada, T. Yamada, L. Q. Phuong, N. Maruyama, H. Nishimura, A. Wakamiya, Y. Murata, and Y. Kanemitsu, J. Am. Chem. Soc. **137**, 10456 (2015).

[13] T. Yamada, Y. Yamada, Y. Nakaike, A. Wakamiya, and Y. Kanemitsu, Phys. Rev. Appl. **7**, 014001 (2017).

[14] T. Yamada, T. Aharen, and Y. Kanemitsu, Phys. Rev. Lett. **120**, 057404 (2018).

[15] J. M. Frost, Phys. Rev. B **96**, 195202 (2017).

[16] K. Miyata and X-Y. Zhu, Nat. Mater. **17**, 379 (2018).

[17] A. J. Neukirch, W. Nie, J.-C. Blancon, K. Appavoo, H. Tsai, M. Y. Sfeir, C. Katan, L. Pedesseau, J. Even, J. J. Crochet, G. Gupta, A. D. Mohite, and S. Tretiak, Nano Lett. **16**, 3809 (2016).

[18] M. Park, A. J. Neukirch, S. E. Reyes-Lillo, M. Lai, S. R. Ellis, D. Dietze, J. B. Neaton, P. Yang, S. Tretiak, and R. A. Mathies, Nat. Commun. **9**, 2525 (2018).

[19] K. Miyata, T. L. Atallah, and X.-Y. Zhu, Sci. Adv. 3, e1701469 (2017).

[20] A. Miyata, A. Mitioglu, P. Plochocka, O. Portugall, J. T.-W. Wang, S. D. Stranks, H. J. Snaith, and R. J. Nicholas, Nat. Phys. **11**, 582 (2015).

[21] K. Galkowski, A. Mitioglu, A. Miyata, P. Plochocka, O. Portugall, G. E. Eperon, J. T.-W. Wang, T. Stergiopoulos, S. D. Stranks, H. J. Snaith, and R. J. Nicholas, Energy Environ. Sci. **9**, 962 (2016).

[22] G. Behnke, H. Buttner and J. Pollmann, Solid State Commun. **20**, 873 (1978).

[23] See Supplemental Material at http://link.aps.org/supplemental/, which includes Refs. [24–34], for details of the sample preparation, experimental setup, theoretical calculation based on Haken model, effective LO phonon energy, and miscellaneous data on magnetoreflectance.

[24] Y. Yamada, T. Yamada, Y. Kanemitsu, Bull. Chem. Soc. Jpn. **90**, 1129 (2017).

[25] R. S. Knox, Solid State Physics, eds. F. Seits and D. Turnbull, (Acadmic Press, 1963) ch. II-6.

[26] K. Kheng, R. T. Cox, Merle Y. d' Aubigné, Franck Bassani, K. Saminadayar, and S. Tatarenko, Phys. Rev. Lett. **71**, 1752 (1993).

[27] G. V. Astakhov, D. R. Yakovlev, V. P. Kochereshko, W. Ossau, J. Nürnberger, W. Faschinger, and G. Landwehr, Phys. Rev. B **60**, R8485 (1999).

[28] C. Klingshirn, Semiconductor Optics, 2nd ed. (Springer, 2005) ch.9.

[29] Y. Yang, M. Yang, Z. Li, R. Crisp, K. Zhu, and M. C. Beard, J. Phys. Chem. Lett. **6**, 4688 (2015).

[30] H. Haken, Z. Phys. **146**, 527 (1956).

[31] Z. Yu, J Phys. Chem. C **121**, 3156-3160 (2017).

[32] M. Nagai, T. Tomioka, M. Ashida, M. Hoyano, R. Akashi, Y. Yamada, T. Aharen, and Y. Kanemitsu, Phys. Rev. Lett. **121**, 145506 (2018).

[33] R. W. Hellwarth and I. Biaggio, Phys. Rev. B **60**, 299 (1999).

[34] M. Bokdam, T. Sander, A. Stroppa, S. Picozzi, D. D. Sarma, C. Franchini, and G. Kresse, Sci. Rep. **6**, 28618 (2016).

[35] Z. Yang, A. Surrente, K. Galkowski, N. Bruyant, D. K. Maude, A. A. Haghighirad, H. J. Snaith, P. Plochocka, and R. J. Nicholas, J. Phys. Chem. Lett. **8**, 1851 (2017).

[36] F. M. Peeters and J. T. Devreese, Phys. Rev. B **31**, 3689 (1985).

[37] F. M. Peeters and J. T. Devreese, Phys. Rev. B **34**, 7246 (1986).

[38] J. Röseler, Phys. Stat. Sol. **25**, 311 (1968).

[39] Y. Imanaka, N. Miura, and H. Kukimoto, Phys. Rev. B **49**, 16965 (1994).

[40] M. Schlipf, S. Ponće, and F. Giustino, Phys. Rev. Lett. **121**, 086402 (2018).

[41] J. Tilchin, D. N. Dirin, G. I. Maikov, A. Sashchiuk, M. V. Kovalenko, and E. Lifshitz, ACS Nano **10**, 6363 (2016).

[42] H. Kunugita, Y. Kiyota, Y. Udagawa, Y. Takeoka, Y. Nakamura, J. Sano, T. Matsushita, T. Kondo and K. Ema, Jpn. J. Appl. Phys. **55**, 060304 (2016).

[43] I. Anusca, S. Balčiūnas, P. Gemeiner, Š. Svirskas, M. Sanlialp, G. Lackner, C. Fettkenhauer, J. Belovickis, V. Samulionis, M. Ivanov, B. Dkhil, J. Banys, V. V. Shvartsman, and D. C. Lupascu, Adv. Energy Mater. **7**, 1700600 (2017).

[44] L. M. Herz, Energy Lett. **2**, 1539 (2017).

[45] H. Wei, D. DeSantis, W. Wei, Y. Deng, D. Guo, T. J. Savenije, L. Cao, and J. Huang, Nat. Mater. **16**, 826 (2017).

[46] Y. Kang and S. Han, Phys. Rev. Applied **10**, 044013 (2018).

[47] A. Lacroix, G. Trambly de Laissardière, P. Quémerais, J.-P. Julien, and D. Mayou, Phys. Rev. Lett. **124**, 196601 (2020).

[48] W. P. D. Wong, J. Yin, B. Chaudhary, X. Y. Chin, D. Cortecchia, S.-Z. A. Lo, A. C. Grimsdale, O. F. Mohammed, G. Lanzani, and C. Soci, ACS Materials Lett. **2**, 20 (2020).

[49] Z. Yang, A. Surrente, K. Galkowski, N, Bruvant, D. K. Maude, A. A. Haghighirad, H. J. Snaith, P. Plochocka, and R. J. Nicholas, J. Phys. Chem. Lett. **8**, 1851 (2017).

[50] T. Kimura, K. Matsumori, K. Oto, Y. Kanemitsu, and Y. Yamada, Appl. Phys. Express **14**, 041009 (2021).

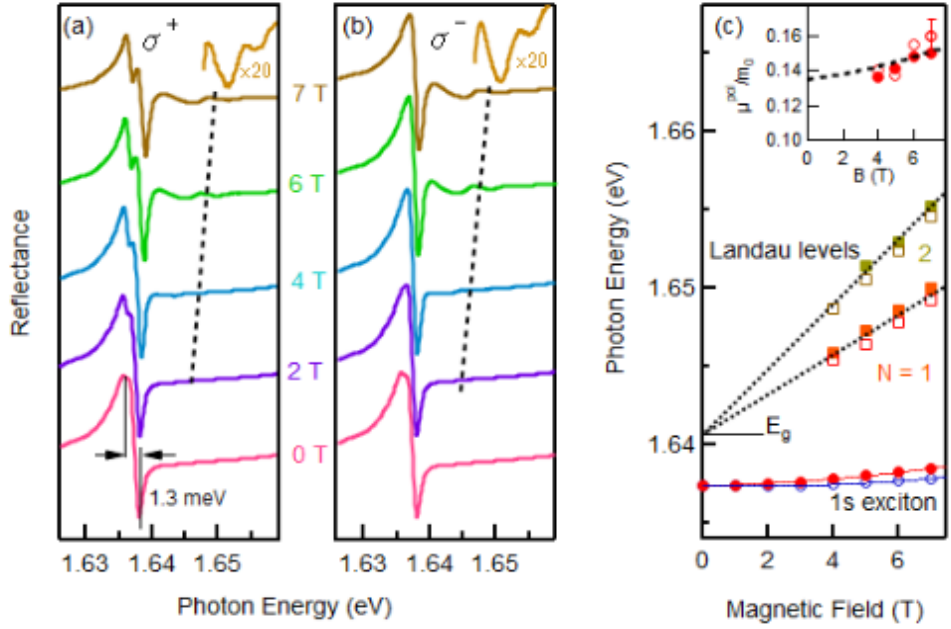

FIG. 1. (Color Online)

Reflectance spectra of $CH_3NH_3PbI_3$ at 1.5 K under different magnetic field for excitation with (a) left and (b) right circular polarized light, denoted by $\sigma^+$ and $\sigma^-$, respectively. The spectra are offset for clarity. The dotted lines indicate the Landau energy for $N = 1$. Magnified views ($\times 20$) of the data obtained under 7 T are also shown. (c) Magnetic-field dependences of the first two Landau levels ($N$ = 1 and 2) and the 1s exciton energy of $CH_3NH_3PbI_3$ under $\sigma^+$ (open symbols) and $\sigma^-$ (filled symbols) excitation conditions. The inset shows the $\mu^{pol}$ as a function of the applied magnetic field for the $\sigma^+$ (open) and $\sigma^-$ (filled) excitation conditions. The dashed curve is the theoretical calculation for $\alpha = 1.4$ from Ref. 37.

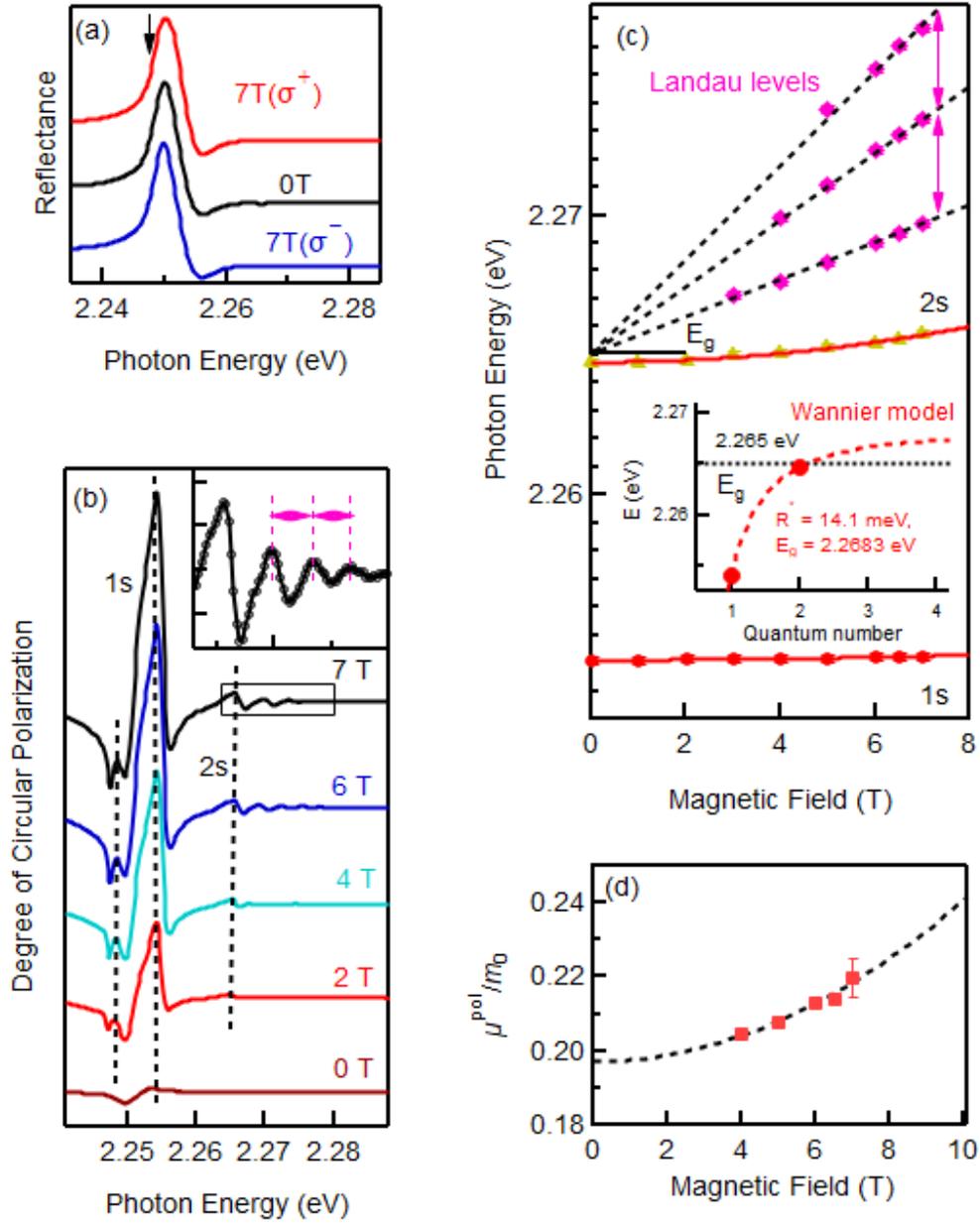

FIG. 2. (Color Online)

(a) Reflectance spectra of CH$_3$NH$_3$PbBr$_3$ at 0 T (black curve) and 7 T for σ$^+$ (red curve) and σ$^-$ excitation (blue curve). The arrow indicates the position of a distinct spectral feature below the 1s exciton energy (see main text). (b) Magnetic-field dependence of the circular polarization spectrum. The broken lines are guides for the eye to clarify the shift directions of spectral features. The curves are offset for clarity. The inset shows an enlarged view of the high-energy region of the spectrum at 7 T. (c) Magnetic-field dependence of the 1s and 2s resonances and Landau levels. The broken lines indicate the linear extrapolations of the Landau levels. The inset shows the exciton resonance energy as a function of quantum number. The broken red curve is the fitting result based on the Wannier model with $R^* = 14.1$ meV and $E_g = 2.268$ eV. (d) The $\mu^{pol}$ as a function of applied magnetic field. The broken curve corresponds to the theoretically predicted result for $\alpha = 2.5$ from Ref. 37.

|  |  |  | $CH_3NH_3PbI_3$ | $CH_3NH_3PbBr_3$ | $CH_3NH_3PbCl_3$ | Refs. |
|---|---|---|---|---|---|---|
| Input parameters | Band mass | $m_e^b/m_0$ | 0.18 | 0.24 | 0.34 | [34] |
|  |  | $m_h^b/m_0$ | 0.25 | 0.30 | 0.35 | [34] |
|  |  | $\mu^b/m_0$ | 0.104 (0.104) | 0.133 (0.117) | 0.172 | [19] [21] |
|  | Phonon energy (meV) | $\hbar\omega_{LO}$ | 16 | 20 | 30 | [32] |
|  | Dielectric constant | $\varepsilon_0$ | 31 | 28 | 15 | [43] |
|  |  | $\varepsilon_\infty$ | 6.8 | 5.2 | 4.2 | [34] |
| Output parameters | Fröhlich constant | $\alpha_e$ | 1.42 | 2.00 | 2.13 |  |
|  |  | $\alpha_h$ | 1.67 | 2.24 | 2.16 |  |
|  | Polaron radius (Å) | $r_e$ | 36 | 28 | 19 |  |
|  |  | $r_h$ | 31 | 25 | 19 |  |
|  | Renormalized mass | $m_e^{pol}/m_0$ | 0.231 | 0.343 | 0.497 |  |
|  |  | $m_h^{pol}/m_0$ | 0.336 | 0.448 | 0.515 |  |
|  |  | $\mu^{pol}/m_0$ | 0.137 (**0.13 ± 0.01**) | 0.194 (**0.20 ± 0.01**) | 0.252 |  |
|  | Rydberg energy (meV) | $R^*$ | 2.43 (**3.1 ± 0.5**) | 10.1 (**11 ± 1**) | 64.3 |  |

Table 1.

Input and output parameters of the calculation. We adopt the literature values of the carrier band masses obtained by theoretical calculation without considering electron-phonon interactions [34], since they are close to the reduced mass estimated from the high-field experiments (given in parentheses [19,21]). The renormalized masses and exciton Rydberg energies determined by the low-field experiment (this work) are also given in parentheses.